\documentclass[twocolumn,aps,prd,showpacs,amsmath,amssymb,nofootinbib,superscriptaddress]{revtex4-1}
\usepackage{graphicx}
\usepackage{dcolumn}
\usepackage{bm}
\usepackage{hyperref}
\usepackage{mathtools}
\usepackage{xcolor}

\begin{document}

\title{Distinguishing wormholes via Einstein rings and global curvature}

%%%%%%%%%%%%%%%%%%%%%%
\author{M. B. Cruz}
\email{messiasdebritocruz@servidor.uepb.edu.br}
\affiliation{Universidade Estadual da Para\'iba (UEPB), \\ Centro de Ci\^encias Exatas e Sociais Aplicadas (CCEA), \\ R. Alfredo Lustosa Cabral, s/n, Salgadinho, Patos - PB, 58706-550 - Brazil.}

%%%%%%%%%%%

\author{C. R. Muniz}
\email{celio.muniz@uece.br}
\affiliation{Universidade Estadual do Cear\'a (UECE), Faculdade de Educa\c{c}\~ao, Ci\^encias e Letras de Iguatu, Av. D\'ario Rabelo s/n, Iguatu - CE, 63.500-00 - Brazil.}
%%%%%%%%%%%

\author{R. M. P. Neves}
\email{raissa.pimentel@uece.br}
\affiliation{Universidade Estadual do Cear\'a (UECE), Faculdade de Educa\c{c}\~ao, Ci\^encias e Letras de Iguatu, Av. D\'ario Rabelo s/n, Iguatu - CE, 63.500-00 - Brazil.}
%%%%%%%%%%%%%%%%%%%%%%%%%%%%%%%%%%%%%%%%%%%%%%%%%%%%%%%%%%%%%%%%%%%%%%
\author{Jonathan A. Rebouças}
\email{jalvesreboucas@ifce.edu.br}
\affiliation{Instituto Federal de Educação Ciências e Tecnologia do Ceará (IFCE), Iguatu, Brazil}
\date{\today}

\begin{abstract}
In this work, we investigate the gravitational lensing properties of a static Ellis-Bronnikov wormhole embedded in a curved Friedmann-Lemaître-Robertson-Walker (FLRW) universe. By employing curvature-dependent cosmological distances, we derive the corresponding weak-field lens equation and demonstrate that the wormhole Einstein ring radius follows a characteristic cubic scaling with cosmological distances, in sharp contrast to the square-root behavior found for Schwarzschild black holes. This distinct scaling leads to a qualitatively different redshift evolution of the 
lensing signal, providing a model-independent geometric diagnostic to discriminate between wormhole and black hole lensing scenarios. Numerical analysis reveals that the interplay between the local wormhole geometry and the FLRW background produces an asymmetric response to spatial curvature that inverts at intermediate redshifts, exhibiting a non-negligible sensitivity even under tight modern constraints such as those from DESI 2024. We also find that Ellis-Bronnikov wormholes are substantially less efficient gravitational lenses than Schwarzschild black holes of comparable physical scale, implying that microarcsecond-scale Einstein rings require macroscopic throat radii. These results suggest that, should a population of cosmological wormholes exist, their lensing signatures could provide a sensitive, complementary 
probe of both exotic spacetime topology and the global geometry of the Universe.
\end{abstract}


\maketitle

%-----------------------------------------------------
%------------------ Introduction ---------------------
%-----------------------------------------------------
\section{Introduction}

Traversable wormholes are exact solutions to Einstein's field equations \cite{EinsteinRosen1935, MorrisThorne1988}, which typically require the presence of exotic matter violating the energy conditions to stabilize their throat \cite{MorrisThorne1988, Visser1989}. These theoretical objects act as bridges connecting distinct regions of spacetime, allowing, in principle, the transit of matter and information. Among the various proposed solutions, the zero-tidal-force class of wormholes, such as the Ellis–Bronnikov model \cite{Ellis1973, Bronnikov1973}, is of particular physical interest. By eliminating destructive tidal forces on a traveler, this class imposes a constant redshift function, usually adopted as $\Phi(r)=0$ in the static reference frame. Recent investigations have revisited the Ellis--Bronnikov geometry in asymptotically safe gravity and have constructed related traversable wormholes in loop quantum gravity \cite{Alencar2021ASG,Cruz2024LQG}. Complementarily, wormhole solutions supported by dark matter in loop quantum cosmology and generalized Ellis--Bronnikov geometries with explicit field sources have enlarged the range of physically motivated configurations available for phenomenological studies \cite{Silva2025DMLQC,Crispim2026GEB}.

The observational study of gravitational lensing is a powerful tool to investigate the spacetime geometry near compact objects. Traditional lensing models typically assume local spherical symmetry and an asymptotically flat background. Although sufficient for most astrophysical scenarios, these assumptions can mask subtle effects arising from exotic topologies when embedded in a global cosmological context. Unlike black holes, whose observational evidence is now firmly established through gravitational wave detections \cite{LIGO_GW150914} and direct horizon-scale imaging \cite{EHT_M87_2019}, wormholes possess a non-trivial spatial structure that alters light propagation, offering a unique arena to test gravitational lensing theory under non-standard boundary conditions.

While the local lensing properties of such objects in asymptotically flat spacetimes are well-established in both weak and strong field regimes \cite{Cramer1995, Perlick2004, Nandi2006}, their observable signatures when embedded in a realistic, expanding Friedmann-Lemaître-Robertson-Walker (FLRW) universe remain under-discussed. Standard lensing studies often approximate the background cosmology as spatially flat ($\Omega_k=0$). However, upcoming deep surveys ($z \gtrsim 1.5$) will be highly sensitive to curvature-induced corrections to angular diameter distances. This observational landscape raises important questions: can the Einstein rings of exotic compact objects serve as a sensitive geometric probe of global spatial curvature $\Omega_k$? Conversely, can the curved cosmological background itself enhance the observational discriminability between wormholes and black holes through their distinct lensing signatures?

In this work, we investigate the optical signatures of a zero-tidal-force wormhole embedded in a curved FLRW cosmological background within the weak-field lensing approximation, building upon the general framework of cosmological wormhole embeddings introduced by Kim \cite{Kim1996}. The choice of a constant redshift function is not only consistent with the original Ellis–Bronnikov geometry, but is also supported by dynamical cosmological wormhole models, where the requirement of vanishing radial energy flux ($T_{tr}=0$) in an expanding universe naturally leads to a constant redshift function \cite{Rahaman:2024bjr}. Related studies of quantum-field effects in FLRW wormhole backgrounds have also shown that locally measured quantities near the throat can depend explicitly on the scale factor and on the global spatial curvature, reinforcing the need to distinguish local wormhole physics from its cosmological embedding \cite{Santos2021Cosmological}.

Accordingly, within this framework, we derive the Einstein ring radius $\theta_E$ in non-flat FLRW geometries and show that the resulting observable $\theta_E(z)$ exhibits a distinctive redshift evolution governed by the global spatial curvature. We demonstrate that the coupling between the local wormhole lensing profile and the cosmological background can break degeneracies with standard compact objects, providing a potentially observable signature of exotic spacetime topology. Finally, we present order-of-magnitude estimates for the expected angular scales and discuss the prospects for detecting such effects with future high-resolution astronomical surveys.

This paper is structured as follows. In Sec. \ref{framework}, we outline the cosmological framework, introducing the metric for a static wormhole embedded in a curved FLRW background, alongside the corresponding null geodesics and conformal time framework. The Sec. \ref{deflec_angle} is dedicated to the evaluation of the deflection angle within the weak-field expansion. In Sec. \ref{cosmo_distance}, we derive the cosmological angular diameter distances, which are subsequently applied in Sec. \ref{Einst_ring} to compute the Einstein ring radius. A comprehensive discussion regarding scaling laws, degeneracy breaking, sensitivity to $\Omega_k$, and order-of-magnitude estimates for detectability is presented in Sec. \ref{discussion}. Finally, our concluding remarks are summarized in Sec. \ref{sec:conclusion}. Throughout this work, we adopt natural units where $c = G = 1$, and the spacetime metric signature is chosen as $(-, +, +, +)$.

%-------------------------------------------------
%-------------- Framework -------------------------
%--------------------------------------------------
\section{Cosmological Framework}
\label{framework}

\subsection{Metric}
We adopt the metric for a wormhole embedded in an FLRW background \cite{Kim1996,Rahaman:2024bjr}. For a zero-tidal-force wormhole ($\Phi(r)=0$, implying $g_{tt}=-1$), the line element is given by:
\begin{equation}
    ds^2 = -dt^2 + a^2(t) \left[ \frac{dr^2}{1 - kr^2 - b(r)/r} + r^2 d\Omega^2 \right],
    \label{eq:metric}
\end{equation}
where $a(t)$ is the cosmic scale factor, $d\Omega^2 = d\theta^2 + \sin^2\theta\, d\phi^2$, and $k\in\{-1,0,+1\}$ denotes the spatial curvature parameter. The shape function $b(r)$ determines the wormhole spatial geometry. For an Ellis--Bronnikov type wormhole \cite{Ellis1973, Bronnikov1973}, this function takes the form:
\begin{equation}
    b(r) = \frac{r_0^2}{r},
    \label{eq:shape}
\end{equation}
where $r_0$ represents the throat radius. The metric \eqref{eq:metric} serves as a phenomenological ansatz assuming the wormhole is comoving with the cosmic expansion, where the global curvature term $kr^2$ couples linearly to the localized spatial geometry of the throat. While a fully self-consistent solution would require solving the Einstein field equations with a combined matter-energy source, this description remains highly accurate as a localized approximation valid for physical scales much smaller than the Hubble radius.

\subsection{Null Geodesics and Conformal Time}
Light propagation is governed by null geodesics ($ds^2=0$). By introducing the conformal time $\eta$, defined via $dt = a(\eta)d\eta$, the line element \eqref{eq:metric} can be rewritten as:
\begin{equation}
    ds^2 = a^2(\eta)\left[-d\eta^2 + \gamma_{ij}dx^i dx^j\right],
\end{equation}
where $\gamma_{ij}$ represents the spatial metric components enclosed in the brackets of Eq.~\eqref{eq:metric}. Due to the conformal invariance of null geodesics, the geometric trajectories of light rays are independent of the global cosmic scaling factor $a(\eta)$ and are determined strictly by the spatial geometry $\gamma_{ij}$ \cite{Perlick2004, Tsukamoto2012}. Consequently, within the localized weak-field lensing approximation, the local deflection angle $\hat{\alpha}$ can be evaluated using the spatial slice of the metric, whereas the global cosmological expansion enters exclusively through the curvature-dependent angular diameter distances within the lens mapping equations.

%-------------------------------------------------
%-------------- Deflection angle ------------------
%--------------------------------------------------
\section{Deflection Angle}
\label{deflec_angle}

We consider the photon motion in the equatorial plane ($\theta=\pi/2$). From the conformal spatial geometry, the radial geodesic equation governing the null trajectory can be expressed as:
\begin{equation}
    \left( \frac{dr}{d\phi} \right)^2 = \frac{r^4}{\xi^2} \left( 1 - kr^2 - \frac{r_0^2}{r^2} \right) \left(1 - \frac{\xi^2}{r^2} \right),
    \label{eq:orbit}
\end{equation}
where $\xi$ is the impact parameter defined by the conserved quantities of the null motion. Consequently, the formal expression for the deflection angle $\hat{\alpha}(\xi)$ is given by:
\begin{equation}
    \hat{\alpha}(\xi) = 2\int_{r_{\min}}^{\infty} \frac{dr}{r\sqrt{\left(1 - kr^2 - \frac{r_0^2}{r^2}\right) \left(\frac{r^2}{\xi^2} - 1\right)}} - \pi,
    \label{eq:alpha_def}
\end{equation}
where $r_{\min}$ denotes the distance of closest approach. For a traversable wormhole, the trajectory never crosses the throat, implying $r_{\min} > r_0$.

%----------------- Weak Field ---------------------
\subsection{Weak-Field Expansion}

We evaluate the integrand to leading order in $r_0/\xi$ while carefully isolating the influence of the spatial curvature parameter $k$. A potential subtlety arises because the $kr^2$ term in the denominator, though negligible near the lens, grows at large distances and could formally alter the asymptotic behavior of the integral. However, for a localized compact lens, the dominant physical contribution to $\hat{\alpha}$ occurs near the turning point where $r \sim \xi$. 

Assuming the impact parameter is well within the cosmological horizon ($\xi \ll H_0^{-1}$), the localized curvature scale satisfies $k\xi^2 \ll 1$. For instance, even for typical galactic scale impact parameters of the order of a parsec, one has $k\xi^2 \sim 10^{-20}$ assuming a curvature radius comparable to the Hubble length. Therefore, the curvature term can be safely neglected within the local deflection zone, recovering the standard weak-field Ellis--Bronnikov deflection. Setting $k=0$ inside Eq.~\eqref{eq:alpha_def}, the turning point reduces to:
\begin{equation}
    r_{\min} = \frac{\sqrt{\xi^2 + \sqrt{\xi^4 + 4r_0^2}}}{\sqrt{2}} \approx \xi ,
\end{equation}
and the leading-order weak-field deflection angle yields \cite{Tsukamoto2012}:
\begin{equation}
    \hat{\alpha}(\xi) \approx \frac{\pi}{4} \left( \frac{r_0}{\xi} \right)^2 + \mathcal{O}\left(\frac{r_0^4}{\xi^4}\right).
    \label{eq:alpha_correct}
\end{equation}

The Eq. \eqref{eq:alpha_correct} describes the local weak-field deflection, where FLRW spatial curvature effects are negligible on the scale of the lensing zone. The global spatial curvature is subsequently incorporated via the non-flat FLRW angular diameter distances that govern light propagation across cosmological baselines.

For comparison, a Schwarzschild black hole of mass $M$ produces a weak-field deflection of $\hat{\alpha}_{\text{BH}}(\xi) \approx 4M/\xi$. This fundamental difference in the scaling law ($1/\xi^2$ versus $1/\xi$) constitutes the core mechanism that drives distinct cosmological evolutions for the resulting gravitational lensing observables.

%--------------------------------------------
%-------- Cosmological Distances ---------------
%-----------------------------------------------
\section{Cosmological Distances}
\label{cosmo_distance}

To describe light propagation across cosmological baselines, we model the background spacetime using the standard late-time $\Lambda$CDM framework. Within the Friedmann--Lema\^{i}tre--Robertson--Walker (FLRW) geometry, the expansion rate is governed by the dimensionless Hubble function $E(z) \equiv H(z)/H_0$, defined as \cite{Weinberg2008}:
\begin{equation}
    E(z) = \sqrt{\Omega_m(1+z)^3 + \Omega_k(1+z)^2 + \Omega_\Lambda} \ ,
\end{equation}
where $\Omega_m$ and $\Omega_\Lambda$ are the present-day matter and dark energy density parameters, respectively, and $\Omega_k = 1 - \Omega_m - \Omega_\Lambda$ accounts for the spatial curvature. Since our analysis targets the late-time universe ($z \le 3$), the radiation density parameter is safely neglected, as its contribution remains well below the percent level throughout the relevant redshift range. For the baseline cosmological parameters, we adopt the recent baryon acoustic oscillation measurements from DESI, which establish the Hubble constant at $H_0 = (67.97 \pm 0.38) \, \text{km} \, \text{s}^{-1} \, \text{Mpc}^{-1}$ \cite{DESI2024VI}.

The comoving radial distance from an observer to an object at redshift $z$ is given by:
\begin{equation}
    \chi(z) = \frac{1}{H_0}\int_0^z \frac{dz'}{E(z')}.
\end{equation}
Accounting for the global spatial geometry dictated by $\Omega_k$, the transverse comoving distance $S_k(\chi)$ takes the standard non-Euclidean form \cite{Hogg, Ferreira_2025}:
\begin{equation}
    S_k(\chi) = \begin{cases}
        \begin{aligned}
            &\frac{1}{H_0\sqrt{|\Omega_k|}}\sin\left(H_0\sqrt{|\Omega_k|}\chi\right), \\ 
            &\quad \text{for } \Omega_k<0 \;\text{(closed)}
        \end{aligned} \\ \\
        \chi, \quad \text{for } \Omega_k=0, \\ \\
        \begin{aligned}
            &\frac{1}{H_0\sqrt{\Omega_k}}\sinh\left(H_0\sqrt{\Omega_k}\chi\right), \\ 
            &\quad \text{for } \Omega_k>0 \;\text{(open)}
        \end{aligned}
    \end{cases}
\end{equation}

In cosmological gravitational lensing, the observational mapping relates angular separations to physical scales via the angular diameter distances. For a lens at redshift $z_L$ and a source at $z_S$, these distances are defined as \cite{Schneider1992}:
\begin{align}
    D_L &= \frac{1}{1+z_L} S_k(\chi(z_L)), \label{eq:DL} \\
    D_S &= \frac{1}{1+z_S} S_k(\chi(z_S)), \\
    D_{LS} &= \frac{1}{1+z_S} S_k\bigl(\chi(z_S)-\chi(z_L)\bigr), \label{eq:DLS}
\end{align}
where we introduced the shorthands $\chi_L \equiv \chi(z_L)$ and $\chi_S \equiv \chi(z_S)$. Crucially, due to the non-Euclidean nature of a curved spatial background ($\Omega_k \neq 0$), the angular diameter distance between the lens and the source is non-additive, meaning $D_{LS} \neq D_S - D_L$. For direct numerical implementation and to circumvent evaluating coordinate differences within the arguments of the transcendental functions, $D_{LS}$ can be recast into a computationally stable form by invoking the subtraction theorems for curvature-dependent functions \cite{Schneider1992}:
\begin{equation}
    \begin{split}
        D_{LS} = \frac{1}{1+z_S} & \left[ S_k(\chi_S)\sqrt{1 + \Omega_k H_0^2 S_k^2(\chi_L)} \right. \\
        & \left. - S_k(\chi_L)\sqrt{1 + \Omega_k H_0^2 S_k^2(\chi_S)} \right].
    \end{split}
    \label{eq:DLS_numerical}
\end{equation}

%----------------------------------------------
%------------ Einstein Ring -----------------------
%-------------------------------------------------
\section{Einstein Ring}
\label{Einst_ring}

The gravitational lensing effect can be analyzed within the standard thin-lens approximation. For a localized lens system, the angular position of the source $\beta$ and the corresponding observed image position $\theta_E$ are coupled through the cosmic distance mapping via \cite{Schneider1992, Congdon}:
\begin{equation}
    \beta = \theta_E - \frac{D_{LS}}{D_S}\hat{\alpha}(\xi),
\end{equation}
where $\hat{\alpha}(\xi)$ represents the physical deflection angle. It is worth noting that while spatial curvature ($\Omega_k \neq 0$) breaks the additivity of angular diameter distances, the structure of the thin-lens equation remains preserved since the non-Euclidean corrections are directly encapsulated within the definition of the geometric distance ratio $D_{LS}/D_S$ \cite{Saha:2024axf}. For an almost perfect alignment between the observer, lens, and source ($\beta \approx 0$), the lens equation simplifies to:
\begin{equation}
    \theta_E = \frac{D_{LS}}{D_S}\,\hat{\alpha}(\xi), \qquad \xi = \theta_E D_L.
\end{equation}
Although perfect alignment is a rare occurrence in observational astrophysics, the Einstein radius $\theta_E$ derived under this condition establishes the characteristic angular scale of the system, serving as a powerful structural parameter even in configurations with broken symmetry.

By substituting the leading-order weak-field deflection for the Ellis--Bronnikov wormhole given in Eq.~\eqref{eq:alpha_correct} into the alignment condition, we obtain:
\begin{equation}
    \theta_E = \frac{D_{LS}}{D_S}\cdot\frac{\pi}{4}\left(\frac{r_0}{\theta_E D_L}\right)^2.
\end{equation}
Multiplying both sides by $\theta_E^2$ and solving for the angular radius yields the expression:
\begin{equation}
    \theta_E^{\text{WH}} = \left( \frac{\pi r_0^2}{4}\, \frac{D_{LS}}{D_S D_L^2} \right)^{1/3}. \label{eq:theta_final}
\end{equation}
For a Schwarzschild black hole of mass $M$, the analogous lensing scale evaluates to the classical relation:
\begin{equation} \label{eq:theta_final_bh}
    \theta_E^{\text{BH}} = \left(\frac{4M D_{LS}}{D_S D_L} \right)^{1/2}.
\end{equation}
The distinct algebraic powers ($1/3$ versus $1/2$) governing the geometric factors in the denominators induce radically different evolutionary patterns across cosmological timescales. This structural discrepancy provides a clean, model-independent observational window to discriminate between the signatures of these two classes of compact objects.

\begin{figure}[htbp]
    \centering
    \includegraphics[width=\columnwidth]{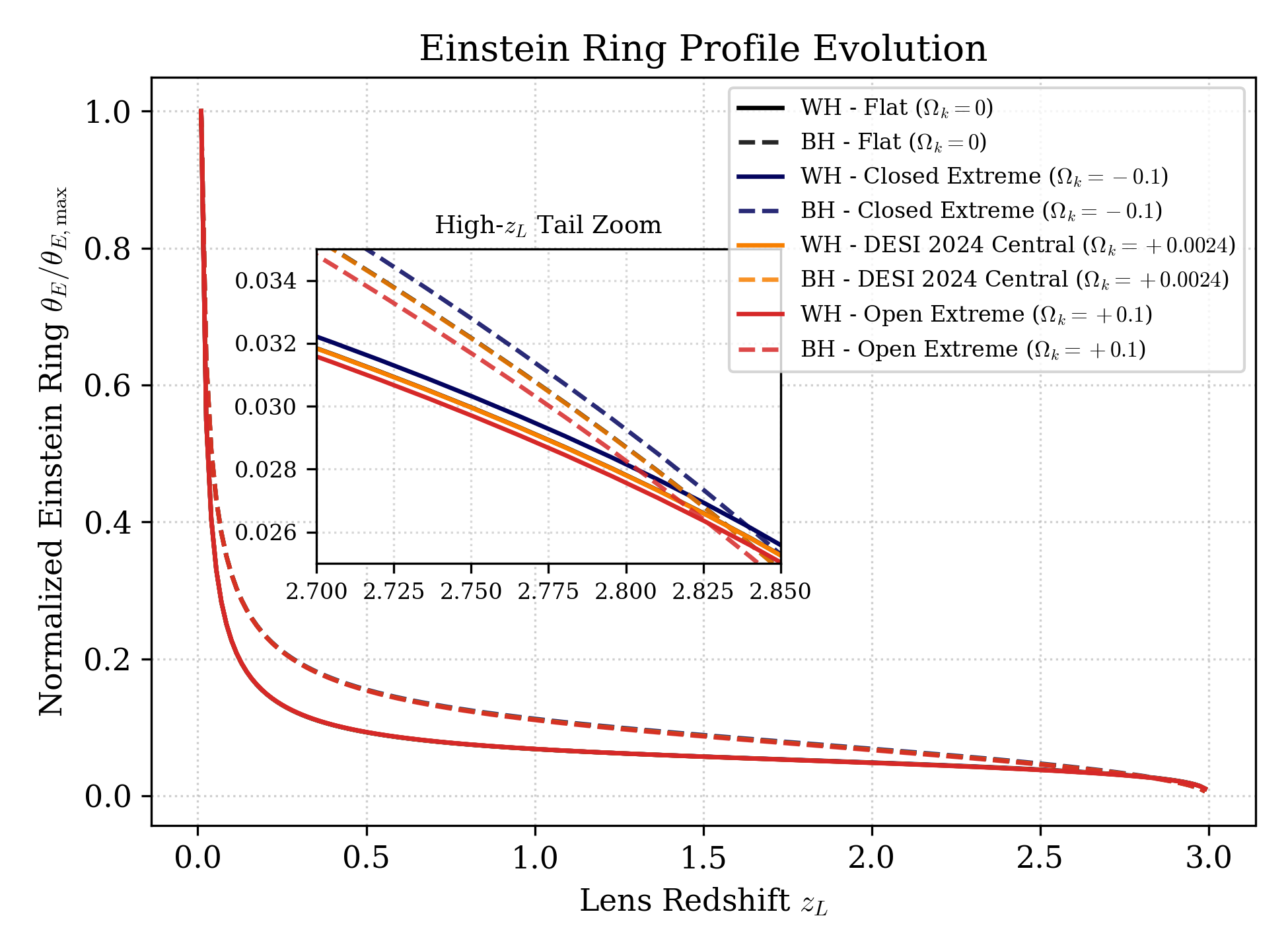}
    \caption{Absolute kinematic profiles: Normalized Einstein ring radius $\theta_E/\theta_{E,\mathrm{max}}$ as a function of the lens redshift $z_L$ for a zero-tidal-force wormhole (solid lines) and a companion Schwarzschild black hole (dashed lines) across different global background spatial geometries. The curves track flat ($\Omega_k=0$, black), closed extreme ($\Omega_k=-0.1$, navy blue), open extreme ($\Omega_k=+0.1$, red), and the observationally constrained DESI 2024 Central ($\Omega_k=+0.0024$, orange dotted) models. The inset magnifies the high-redshift tail ($2.70 \le z_L \le 2.85$) to highlight how the geometric split lifts the metric degeneracy near the source.}
    \label{fig:profile_evolution}
\end{figure}

\begin{figure}[htbp]
    \centering
    \includegraphics[width=\columnwidth]{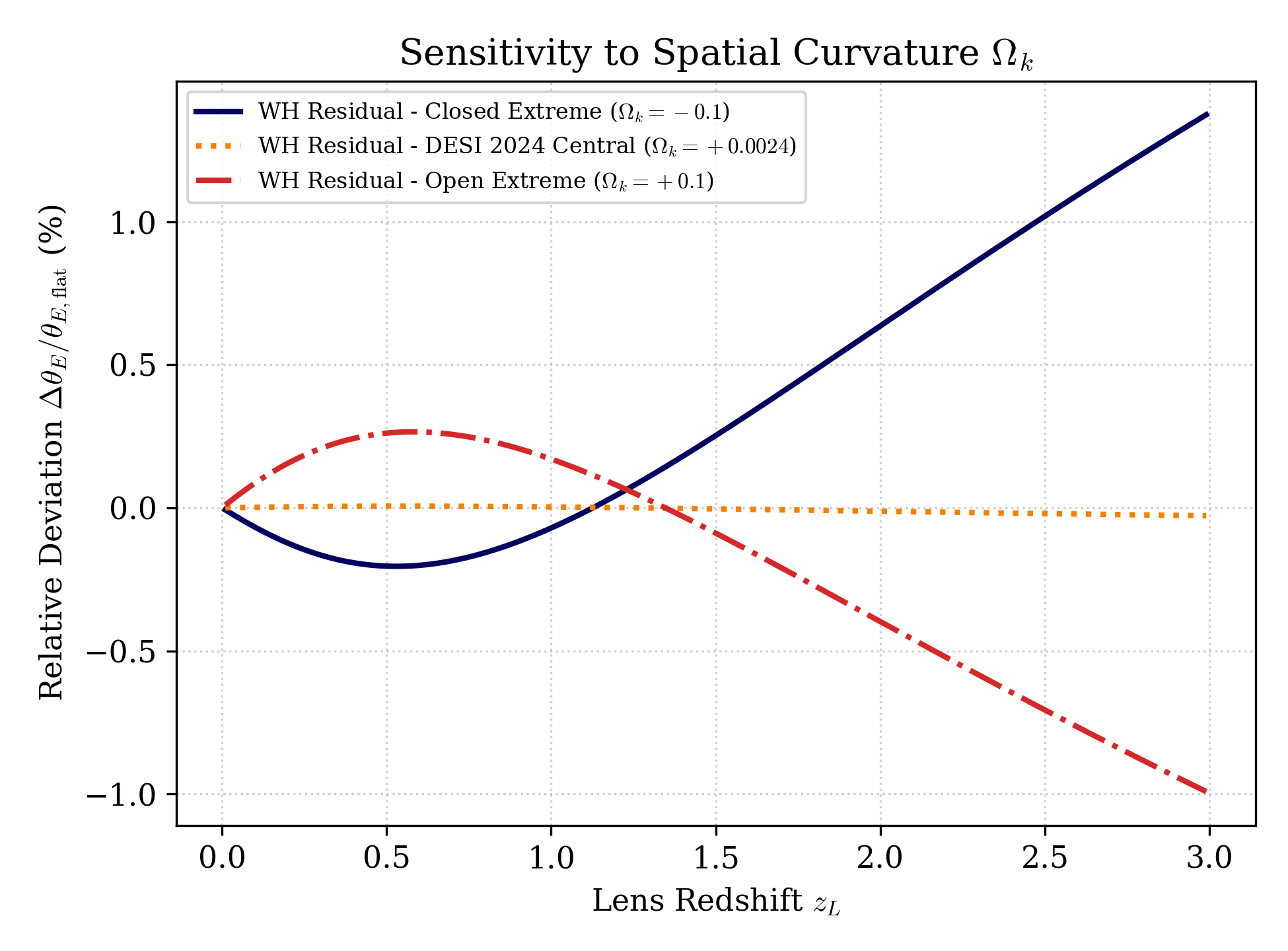}
    \caption{Isolated curvature residuals: Relative deviation $\Delta\theta_E / \theta_{E,\mathrm{flat}}$ (\%) of the wormhole Einstein ring radius as a function of $z_L$. This panel subtracts the flat-space baseline to isolate the exact systematic impact introduced solely by the spatial curvature parameter $\Omega_k$, including the micro-percentage shift caused by the DESI 2024 Central parameter ($\Omega_k = +0.0024$, orange dotted curve).}
    \label{fig:curvature_sensitivity}
\end{figure}

\begin{figure}[htbp]
    \centering
    \includegraphics[width=\columnwidth]{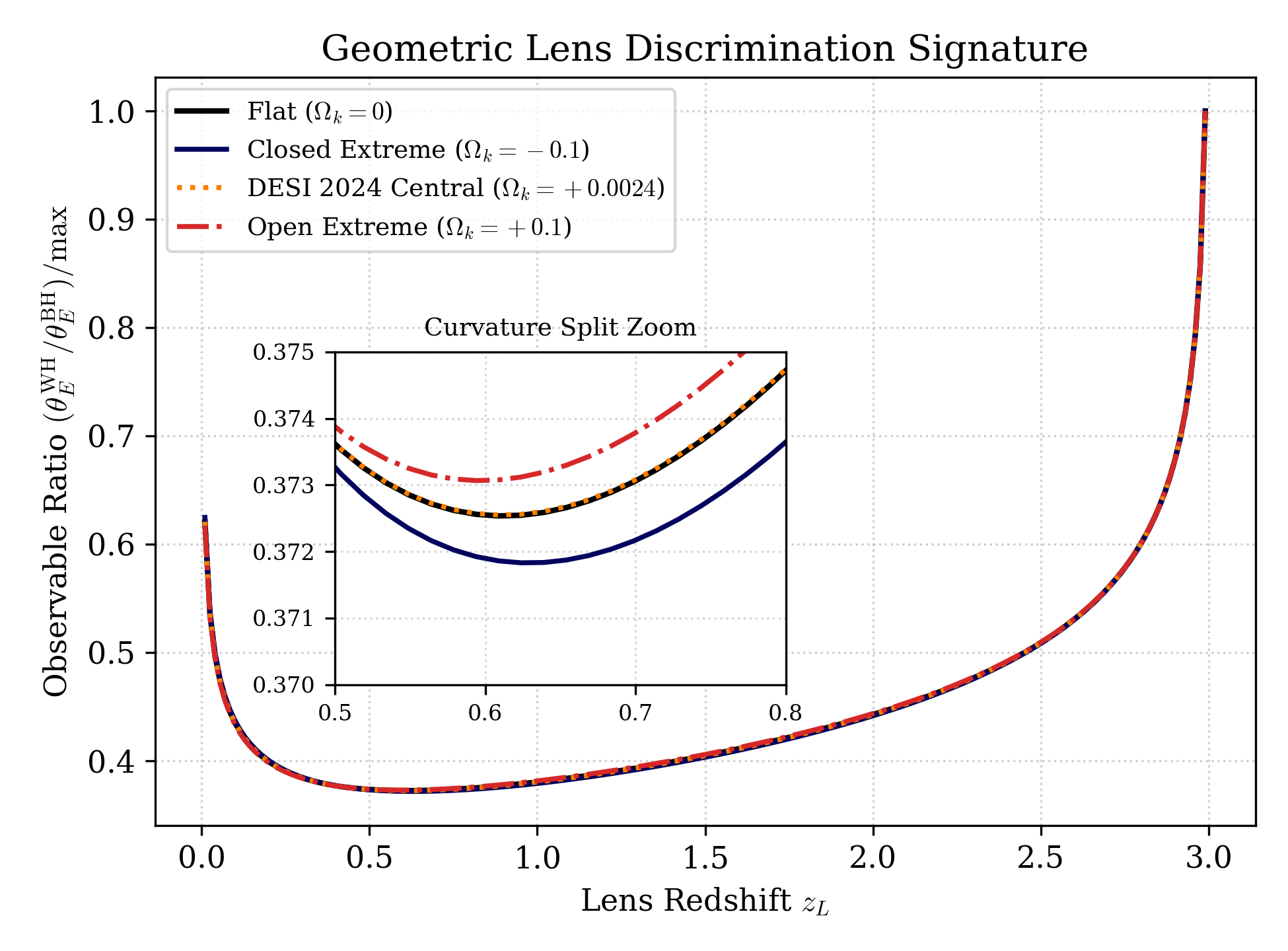}
    \caption{Model-independent lens ratio: The normalized geometric discrimination signature $(\theta_E^{\mathrm{WH}} / \theta_{E}^{\mathrm{BH}}) / \mathrm{max}$ as a function of the lens redshift. By dividing the two radii, local intrinsic parameters ($M$ and $r_0$) cancel out identically, isolating a pure geometric fingerprint. The inset focuses on the Curvature Split Zoom ($0.5 \le z_L \le 0.8$) within the characteristic valley to illustrate how global curvature alters the local discrimination depth.}
    \label{fig:lens_discrimination}
\end{figure}

\begin{figure}[htbp]
    \centering
    \includegraphics[width=\columnwidth]{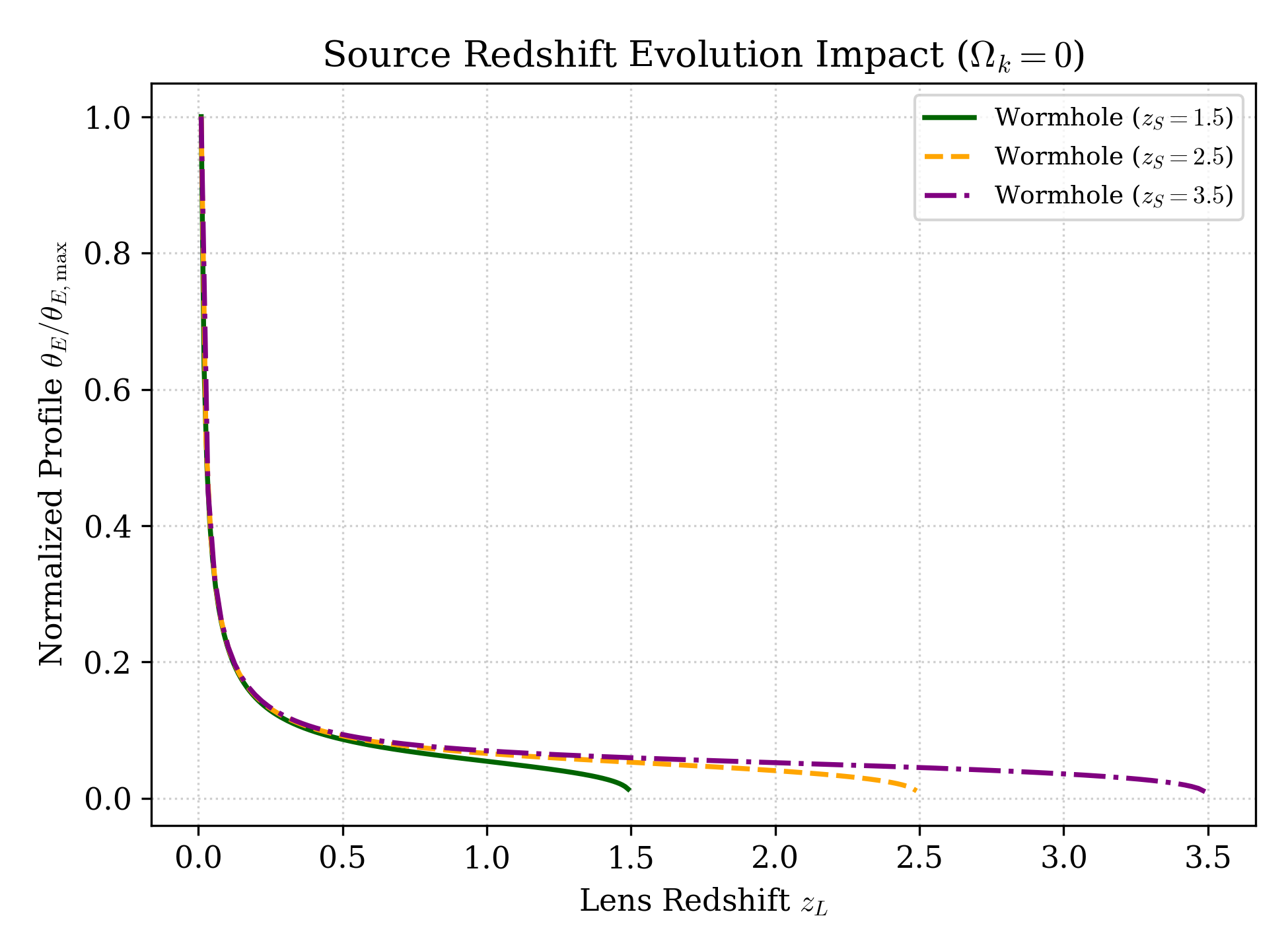}
    \caption{Parametric source boundaries: Normalized wormhole Einstein ring profiles evaluated under varying source plane assignments ($z_S = 1.5$, dark green solid; $z_S = 2.5$, orange dashed; $z_S = 3.5$, purple dash-dotted) within a strictly flat spatial geometry ($\Omega_k = 0$), mapping how the source boundary controls the kinematic range.}
    \label{fig:source_redshift}
\end{figure}

To quantitatively evaluate these lensing expressions under realistic cosmic backgrounds, we perform numerical simulations by fixing a baseline source redshift at $z_S = 3.0$ (except where parametrically varied). For the background parameters, we adopt a standard representative baseline setting $\Omega_m = 0.3$ and $H_0 = 70 \, \text{km} \, \text{s}^{-1} \, \text{Mpc}^{-1}$, which closely benchmarks and remains highly compatible with the recent precision cosmological constraints from DESI 2024 data \cite{DESI2024VI}. The results of these numerical integrations are organized and analyzed in Figs.~\ref{fig:profile_evolution} to \ref{fig:source_redshift}.
In Fig.~\ref{fig:profile_evolution}, we present the primary evolution of the normalized Einstein ring profiles, capturing the absolute kinematic scale of the lensing systems. Near the observer ($z_L \to 0$), both lens profiles exhibit a characteristic geometric divergence driven by the angular diameter distance mapping. However, as $z_L$ increases, the wormhole radius (solid lines) undergoes a much steeper attenuation compared to the black hole counterpart (dashed lines). This accelerated attenuation is a direct consequence of the stronger distance dependency ($\propto D_L^{-2/3}$ inside the cubic root), demonstrating that the wormhole lensing scale is heavily suppressed for lenses positioned at cosmological depths. On this macroscale, the subtle realistic curvature contribution from the DESI 2024 data appears nearly degenerate with the flat space baseline, requiring a dedicated residual inspection.

To map the precise role played by global geometry on the lensing signature without the dominance of the absolute kinematic decay, Fig.~\ref{fig:curvature_sensitivity} tracks the relative residual deviation:
\begin{equation}
    \frac{\Delta\theta_E}{\theta_{E,\mathrm{flat}}} = 100 \times \frac{(\theta_E - \theta_{E,\mathrm{flat}})}{\theta_{E,\mathrm{flat}}},
\end{equation}
exclusively for the wormhole case. The cosmological background introduces a severe asymmetry that inverts at $z_L \approx 1.3$, marking the transition between local expansion rate dynamics and global spatial volume integration. At high redshifts ($z_L > 1.5$), a closed universe ($\Omega_k < 0$, navy blue line) bounds the distance mapping through trigonometric functions, triggering a positive deviation up to $\sim +1.4\%$. Conversely, open universes ($\Omega_k > 0$) broaden the comoving distance via hyperbolic functions, depressing the geometric efficiency and causing a negative residual tail. Crucially, the realistic cosmic background constrained by the DESI 2024 central value ($\Omega_k = +0.0024$, orange dotted line) imposes a subtle but well-behaved micro-percentage suppression ($\approx -0.03\%$) near the source threshold, proving that the underlying formalism is highly sensitive to modern precision cosmology parameters.

A major challenge in exotic lensing phenomenology is breaking the parameter degeneracy associated with the unconstrained mass and throat scales ($M$ and $r_0$). We address this constraint in Fig.~\ref{fig:lens_discrimination} by constructing the geometric discrimination ratio $(\theta_E^{\mathrm{WH}} / \theta_{E}^{\mathrm{BH}}) / \mathrm{max}$, which completely decouples the intrinsic scales of the compact objects from the global metric background. Interestingly, while the curves converge to unity at the boundaries ($z_L \to 0$ and $z_L \to z_S$) due to the global coordinate limits of the FLRW framework, the intermediate regime reveals a clear, model-independent separation. The position and depth of the local minimum shift deterministically with $\Omega_k$, dropping to a characteristic valley near $\sim 0.372$. As exposed by the inner zoom panel, the structural stability of this ratio remains robust even under conservative, observationally viable curvature parameters, providing an elegant method to discriminate the lens nature while cross-checking the background geometry.

Finally, Fig.~\ref{fig:source_redshift} tests the structural stability of our modeling by varying the source plane position ($z_S = 1.5, 2.5, 3.5$) in a controlled flat spatial geometry ($\Omega_k = 0$). This operational setup ensures that the observed variations are driven exclusively by the choice of the background source plane rather than curvature changes. The normalized profiles demonstrate a remarkable scale invariance at low lens redshifts ($z_L < 0.5$), where the lines overlap seamlessly. This behavior confirms that the sharp near-field decay is an intrinsic property of the wormhole space-time geometry, remaining entirely uncoupled from the depth of the background source. Conversely, deeper sources ($z_S = 3.5$) widen the optical efficiency window, showing that high-redshift surveys maximize the observational tracking of the lens signatures before the final local kinematic collapse ($z_L \to z_S$) takes place.

%--------------------------------------------------
%------------ Discussion and Prospects ------------
%--------------------------------------------------
\section{Discussion and Observational Prospects}
\label{discussion}

%--------------------------------------------------
\subsection{Scaling Laws and Degeneracy Breaking}

The underlying space-time geometry of the lens imprints a distinct signature on the angular scale of the system. The Eq.~\eqref{eq:theta_final} establishes that the wormhole Einstein radius scales as $\theta_E^{\text{WH}} \propto D_{\text{eff}}^{1/3}$, where the effective distance is defined as $D_{\text{eff}} = D_{LS}/(D_S D_L^2)$. In contrast, a standard Schwarzschild black hole follows the classical scaling law $\theta_E^{\text{BH}} \propto [D_{LS}/(D_S D_L)]^{1/2}$. 

As numerical results in Fig.~\ref{fig:profile_evolution} demonstrate, the higher power of the lens distance in the denominator for the wormhole profile ($D_L^2$ versus $D_L$) induces a significantly sharper attenuation as the lens shifts away from the observer ($z_L \to z_S$). Conversely, the cubic root acting upon $D_{\text{eff}}$ flattens the profile at intermediate regimes. Therefore, a statistical population of lensed systems displaying a steep early-decay profile followed by a shallow redshift dependence at higher lens positions would characteristically favor a wormhole morphology over a standard black hole interpretation.

%---------------------------------------------------
\subsection{Sensitivity to $\Omega_k$ and Cosmic Residuals}

Since the angular diameter distances $D_L$, $D_S$, and $D_{LS}$ are coupled to the spatial curvature through the curvature-dependent comoving distance function $S_k(\chi)$, the geometric factor $D_{\text{eff}}$ inherits global topological corrections. Our numerical analysis reveals a major, non-trivial asymmetry in how global geometry shifts the lensing scale, triggering an inversion signature at $z_L \approx 1.3$.

As quantified by the relative residuals in Fig.~\ref{fig:curvature_sensitivity}, for high-redshift thresholds ($z_L > 1.5$), a closed universe ($\Omega_k < 0$) focuses light more efficiently due to trigonometric sinusoidal mapping ($\sin$), triggering a positive deviation that reaches up to $\sim +1.4\%$ for the extreme model ($\Omega_k = -0.1$). Conversely, open spatial geometries ($\Omega_k > 0$) broaden the comoving distance via hyperbolic functions ($\sinh$), causing a systematic suppression of the order of $\sim -1.0\%$ for $\Omega_k = +0.1$. 

Crucially, under the realistic cosmic background constrained by the latest DESI 2024 central value ($\Omega_k = +0.0024$), our framework tracks a well-behaved micro-percentage suppression of approximately $-0.03\%$ near the source threshold. Although these precise observationally-bounded curvature variations remain subtle, next-generation Very Long Baseline Interferometry configurations—such as the next-generation Event Horizon Telescope (ngEHT) \cite{Johnson2023}—aim for an angular resolution scale of $\sim 1\,\mu\text{as}$. This unprecedented resolution makes these high-$z_L$ metric splitting features key targets for precision lensing constraints.

%---------------------------------------------------
\subsection{Order-of-Magnitude Estimates and Detectability}

To assess the observational feasibility of the proposed signature, we estimate the absolute angular scale of the wormhole Einstein ring using the expression:
\begin{equation}
\theta_E^{\rm WH}
=
\left(
\frac{\pi r_0^2}{4}
\frac{D_{LS}}{D_S D_L^2}
\right)^{1/3}
\equiv
\left(
\frac{\pi r_0^2}{4}D_{\rm eff}
\right)^{1/3}.
\end{equation}
For a representative flat cosmology with $\Omega_m=0.3$, $H_0=70\,\text{km} \, \text{s}^{-1} \, \text{Mpc}^{-1}$, $z_S=3$, and $z_L=1$, one finds:
\begin{equation}
D_{\rm eff}
=
\frac{D_{LS}}{D_SD_L^2}
\simeq
1.8 \times 10^{-52}\,\text{m}^{-2}.
\end{equation}
This gives the useful numerical estimate:
\begin{equation}
\theta_E^{\rm WH}
\simeq
2.3 \times 10^{-4}\,\mu\text{as}
\left(
\frac{r_0}{3\,\text{km}}
\right)^{2/3}
\left(
\frac{D_{\rm eff}}{1.8 \times 10^{-52}\,\text{m}^{-2}}
\right)^{1/3}.
\end{equation}

Table~\ref{tab:detectability} highlights the striking difference in lensing efficiency between Ellis--Bronnikov wormholes and Schwarzschild black holes of comparable physical scale. For $r_0\simeq3\,\text{km}$, the predicted wormhole Einstein ring is only $\theta_E^{\rm WH}\simeq2.3 \times 10^{-4}\,\mu\text{as}$, whereas a $1\,M_\odot$ Schwarzschild black hole produces $\theta_E^{\rm BH}\simeq1.55\,\mu\text{as}$. More generally, black holes generate Einstein rings that are typically three to four orders of magnitude larger than those of Ellis--Bronnikov wormholes, reflecting the different weak-field scaling laws $\hat{\alpha}_{\rm BH}\propto\xi^{-1}$ and $\hat{\alpha}_{\rm WH}\propto\xi^{-2}$.

%---------------------------------------------------
\begin{table}[htbp]
\centering
\caption{Comparison of Einstein ring sizes for Ellis--Bronnikov wormholes and Schwarzschild black holes with the same characteristic length scale. We assume $z_L=1$, $z_S=3$, $\Omega_m=0.3$, and $H_0=70\,\text{km} \, \text{s}^{-1} \, \text{Mpc}^{-1}$.}
\begin{tabular}{lccc}
\hline
Scale & WH ($\mu\text{as}$) & BH ($\mu\text{as}$) & BH/WH \\
\hline
$3\,\text{km}$ & $2.25 \times 10^{-4}$ & $1.55$ & $6.88 \times 10^{3}$ \\
$30\,\text{km}$ & $1.05 \times 10^{-3}$ & $4.90$ & $4.69 \times 10^{3}$ \\
$R_\odot$ & $0.851$ & $7.47 \times 10^{2}$ & $8.77 \times 10^{2}$ \\
$0.1\,\text{AU}$ & $6.58$ & $3.46 \times 10^{3}$ & $5.26 \times 10^{2}$ \\
$1\,\text{AU}$ & $30.55$ & $1.10 \times 10^{4}$ & $3.58 \times 10^{2}$ \\
\hline
\end{tabular}
\label{tab:detectability}
\end{table}
%---------------------------------------------------

Consequently, detectable wormhole lensing requires substantially larger throats. While stellar-scale throats remain far below current and foreseeable observational capabilities, solar-radius-scale throats approach the microarcsecond regime, and AU-scale wormholes produce Einstein rings of several tens of microarcseconds. These angular scales are comparable to the target resolution of current and next-generation very-long-baseline interferometric facilities such as the EHT and ngEHT \cite{Johnson2023}. By contrast, optical and near-infrared surveys such as \textit{Euclid} operate at angular resolutions of order $0.1''$ \cite{Euclid2011,Euclid2023}, several orders of magnitude above the scales predicted here.

%---------------------------------------------------
\subsection{Geometric Discrimination Through Redshift Evolution}

The different distance scalings in Eqs.~\eqref{eq:theta_final} and \eqref{eq:theta_final_bh} motivate a direct comparison between wormhole and black hole Einstein radii as functions of the lens redshift. The Fig.~\ref{fig:lens_discrimination} shows the normalized ratio $\theta_E^{\rm WH}(z_L)/\theta_E^{\rm BH}(z_L)$ across different spatial curvatures. Although its overall normalization depends on the throat radius $r_0$ and the black hole mass $M$, its redshift evolution reflects the distinct geometric combinations $D_{LS}/(D_S D_L^2)$ and $D_{LS}/(D_S D_L) $ entering the two lensing models. 

As revealed by the characteristic profile valley in Fig.~\ref{fig:lens_discrimination}, the geometric fingerprint drops to a deterministic minimum of $\sim 0.372$ around $z_L \approx 0.6$. The inner split zoom panel confirms that while the curves merge at the global boundaries due to FLRW metric dominance, the intermediate depth and shifting of this valley depend strictly on $\Omega_k$, providing a robust diagnostic to simultaneously determine the lens nature and probe the geometric background.

Furthermore, Fig.~\ref{fig:source_redshift} shows that the normalized wormhole profiles under varying source plane locations ($z_S = 1.5, 2.5, 3.5$) preserve a strict scale invariance in the low-redshift regime ($z_L < 0.5$). This confirms that the qualitative features and the sharp near-field decay discussed above are governed primarily by the intrinsic lens geometry and the background cosmology rather than by specific source configurations, guaranteeing the phenomenological robustness of the signatures.

%---------------------------------------------------
\subsection{Robustness Against Cosmological Parameter Uncertainties}

An important question is whether the geometric signatures discussed above remain sensitive to the wormhole geometry when uncertainties in the background cosmological parameters are taken into account. This issue is particularly relevant in light of the current Hubble-tension scenario \cite{DiValentino:2021izs,Riess:2021jrx}.

At the analytical level, the dependence on the Hubble constant can be estimated directly from the scaling laws derived in Sec.~\ref{discussion}. Since the angular-diameter distances scale as $D\propto H_0^{-1}$ in the standard FLRW framework, one finds $\theta_E^{\rm WH}\propto H_0^{2/3}$ for the Ellis--Bronnikov wormhole and $\theta_E^{\rm BH}\propto H_0^{1/2}$ for the Schwarzschild case. Consequently, the ratio $R\equiv\theta_E^{\rm WH}/\theta_E^{\rm BH}$ depends only weakly on the Hubble constant, $R\propto H_0^{1/6}$.

This weak scaling implies that even relatively large variations in $H_0$ translate into comparatively small changes in the discrimination ratio. For example, a $10\%$ shift in the Hubble constant modifies $R$ by only about $1.6\%$. A similar partial cancellation occurs for the matter-density parameter through the distance ratios entering the lens equation. Therefore, the qualitative distinction between the wormhole and black hole scaling laws is expected to be considerably less sensitive to cosmological parameter uncertainties than the absolute Einstein-ring radii themselves.

%--------------------------------------------------
%----------------- Conclusions --------------------
%--------------------------------------------------
\section{Conclusion}
\label{sec:conclusion}

In this work, we have investigated the Einstein ring radius generated by an Ellis--Bronnikov wormhole embedded in a curved FLRW cosmological background. We have shown that the resulting lensing signal differs fundamentally from the standard Schwarzschild black hole case, both through its characteristic weak-field power-law behavior and through its distinct coupling to the cosmological distance factors. As a consequence, wormhole and black hole lenses exhibit markedly different redshift evolutions, providing a robust mathematical and phenomenological avenue for their observational discrimination.

Our numerical simulations demonstrate that the global spatial curvature parameter $\Omega_k$ imprints a subtle, asymmetric modulation on the wormhole lensing signal. Contrary to local expectations, at high lens redshifts, closed geometries trigger a relative enhancement of the Einstein ring radius due to trigonometric focusing effects, whereas open universes introduce an asymptotic suppression tail. Crucially, our framework proves sensitive enough to track the micro-percentage corrections imposed by the realistic DESI 2024 central constraints ($\Omega_k = +0.0024$) \cite{DESI2024VI}. We have further shown that the redshift evolution of the ratio $\theta_E^{\rm WH}/\theta_E^{\rm BH}$ eliminates local scale degeneracies, providing a robust, model-independent geometric diagnostic characterized by a deterministic minimum valley near $z_L \approx 0.6$. In addition, the normalized profiles preserve a strict scale invariance against variations in the source redshift at low-field regimes, confirming that the qualitative trends identified herein are governed primarily by the intrinsic lens topology and the global cosmological background.

Order-of-magnitude estimates reveal that Ellis--Bronnikov wormholes are substantially less efficient gravitational lenses than Schwarzschild black holes of comparable physical scale, owing to their steeper deflection scaling ($\hat{\alpha}_{\rm WH}\propto\xi^{-2}$ versus $\hat{\alpha}_{\rm BH}\propto\xi^{-1}$). While kilometer-scale throats generate Einstein rings far below current observational thresholds, solar-scale throats approach the microarcsecond regime, and AU-scale wormholes produce angular scales of several tens of microarcseconds. Such signatures are potentially accessible to current and next-generation VLBI facilities, such as the EHT and ngEHT \cite{Johnson2023}. 

We conclude that, should a population of such exotic compact objects exist in the deep sky, their lensing signatures may provide a complementary probe of both cosmic curvature and exotic spacetime topology. The Ellis geometry offers a particularly clean framework for this purpose, since the absence of a redshift function allows the lensing signal to be traced directly to the throat geometry and the underlying FLRW metric. Finally, while the zero-tidal metric remains a useful phenomenological model, extending the present analysis to fully self-consistent dynamical cosmological wormhole solutions constitutes a natural next step toward establishing a more complete gravitational and structural description of these systems.

%----------------------------------------------------
%-------------- Acknowledgements ---------------
%----------------------------------------------------
\section*{Acknowledgments}
\noindent M. B. Cruz acknowledges financial support from the Conselho Nacional de Desenvolvimento Cient\'{i}fico e Tecnol\'{o}gico (CNPq), through grant 301812/2026-8. CRM would like to thank the Conselho Nacional de Desenvolvimento Cient\'{i}fico e Tecnol\'{o}gico (CNPq) for partial financial support, through grant 301122/2025-3. RMPN acknowledges financial support from the Funda\c{c}\~{a}o Cearense de Apoio ao Desenvolvimento Cient\'{i}fico e Tecnol\'{o}gico (FUNCAP), under grant BP6-0241-00123.01.00/25.

%---------------------------------------------------
%--------- References -----------------------
%-------------------------------------------------


\begin{thebibliography}{99}

\bibitem{EinsteinRosen1935}
A.~Einstein and N.~Rosen,
%``The Particle Problem in the General Theory of Relativity,''
Phys. Rev. \textbf{48} (1935), 73-77.

\bibitem{MorrisThorne1988}
M.~S.~Morris and K.~S.~Thorne,
%``Wormholes in space-time and their use for interstellar travel: A tool for teaching general relativity,''
Am. J. Phys. \textbf{56} (1988), 395-412.

\bibitem{Visser1989}
M.~Visser,
%``Traversable wormholes: Some simple examples,''
Phys. Rev. D \textbf{39} (1989), 3182-3184.

\bibitem{Ellis1973}
H.~G.~Ellis,
%``Ether flow through a drainhole - a particle model in general relativity,''
J. Math. Phys. \textbf{14} (1973), 104-118.

\bibitem{Bronnikov1973}
K.~A.~Bronnikov,
%``Scalar-tensor theory and scalar charge,''
Acta Phys. Polon. B \textbf{4} (1973), 251-266.

\bibitem{Alencar2021ASG}
G.~Alencar, V.~B.~Bezerra, C.~R.~Muniz and H.~S.~Vieira,
% ``Ellis--Bronnikov Wormholes in Asymptotically Safe Gravity,''
Universe \textbf{7} (2021) no.~7, 238.

\bibitem{Cruz2024LQG}
M.~B.~Cruz, R.~M.~P.~Neves and C.~R.~Muniz,
% ``Traversable wormholes from Loop Quantum Gravity,''
J. Cosmol. Astropart. Phys. \textbf{05} (2024), 016.

\bibitem{Silva2025DMLQC}
M.~V.~de~S.~Silva, G.~Alencar, R.~N.~Costa~Filho, R.~M.~P.~Neves and C.~R.~Muniz,
% ``Traversable wormholes sourced by dark matter in loop quantum cosmology,''
Eur. Phys. J. Plus \textbf{140} (2025) no.~4, 289.

\bibitem{Crispim2026GEB}
T.~M.~Crispim, G.~Alencar and C.~R.~Muniz,
% ``Field Sources for Generalized Ellis--Bronnikov Wormhole,''
Class. Quantum Grav. \textbf{43} (2026) no.~11, 115013.

\bibitem{EHT_M87_2019}
K.~Akiyama \textit{et al.} [Event Horizon Telescope],
%``First M87 Event Horizon Telescope Results. I. The Shadow of the Supermassive Black Hole,''
Astrophys. J. Lett. \textbf{875} (2019), L1.

\bibitem{LIGO_GW150914}
B.~P.~Abbott \textit{et al.} [LIGO Scientific and Virgo],
%``Observation of Gravitational Waves from a Binary Black Hole Merger,''
Phys. Rev. Lett. \textbf{116} (2016) no.6, 061102.

\bibitem{Cramer1995}
J.~G.~Cramer, R.~L.~Forward, M.~S.~Morris, M.~Visser, G.~Benford and G.~A.~Landis,
%``Natural wormholes as gravitational lenses,''
Phys. Rev. D \textbf{51} (1995), 3117-3120.

\bibitem{Perlick2004}
V.~Perlick,
%``On the Exact gravitational lens equation in spherically symmetric and static space-times,''
Phys. Rev. D \textbf{69} (2004), 064017.

\bibitem{Nandi2006}
K.~K.~Nandi, Y.~Z.~Zhang and A.~V.~Zakharov,
%``Gravitational lensing by wormholes,''
Phys. Rev. D \textbf{74} (2006), 024020.

\bibitem{Kim1996}
S.~W.~Kim,
%``The Cosmological model with traversable wormhole,''
Phys. Rev. D \textbf{53} (1996), 6889-6892.

\bibitem{Rahaman:2024bjr}
F.~Rahaman and B.~S.~Choudhury,
%``Evolving wormhole geometry from dark matter energy density,''
Eur. Phys. J. C \textbf{84} (2024), 504.

\bibitem{Santos2021Cosmological}
A.~C.~L.~Santos, C.~R.~Muniz and L.~T.~Oliveira,
% ``Casimir effect nearby and through a cosmological wormhole,''
EPL \textbf{135} (2021) no.~1, 19002.

\bibitem{Tsukamoto2012}
N.~Tsukamoto, T.~Harada and K.~Yajima,
%``Can we distinguish between black holes and wormholes by their Einstein ring systems?,''
Phys. Rev. D \textbf{86} (2012), 104062.

\bibitem{Weinberg2008}
S.~Weinberg,
\textit{Cosmology},
Oxford University Press, Oxford (2008).

\bibitem{DESI2024VI}
A.~G.~Adame \textit{et al.} [DESI Collaboration],
%``DESI 2024 VI: Cosmological constraints from the measurements of baryon acoustic oscillations,''
J. Cosmol. Astropart. Phys. \textbf{2025} (2025) no.02, 021.

\bibitem{Hogg}
D.~W.~Hogg,
%``Distance Measures in Cosmology,''
arXiv:astro-ph/9905116 (1999).

\bibitem{Ferreira_2025}
P.~G.~Ferreira and A.~Roskill,
%``A Short Introduction to Cosmology and Its Current Status,''
SciPost Phys. Lect. Notes \textbf{109} (2025).

\bibitem{Schneider1992}
P.~Schneider, J.~Ehlers and E.~E.~Falco,
\textit{Gravitational Lenses},
Springer-Verlag, Berlin Heidelberg (1992).

\bibitem{Congdon}
A.~B.~Congdon and C.~R.~Keeton,
\textit{Principles of Gravitational Lensing},
Springer, Cham (2018).

\bibitem{Saha:2024axf}
P.~Saha \textit{et al.},
%``Fundamentos da Lente Gravitacional Forte,''
Space Sci. Rev. \textbf{220} (2024), 12.

\bibitem{Johnson2023}
M.~D.~Johnson \textit{et al.},
%``Key Science Goals for the Next-Generation Event Horizon Telescope,''
Galaxies \textbf{11} (2023) no.3, 61.

\bibitem{Euclid2011}
R.~Laureijs \textit{et al.} [Euclid Collaboration],
%``Euclid Definition Study Report,''
arXiv:1110.3193 [astro-ph.CO] (2011).

\bibitem{Euclid2023}
S.~Casas \textit{et al.} [Euclid Collaboration],
%``Euclid: Constraints on f(R) cosmologies from the spectroscopic and photometric primary probes,''
Astron. Astrophys. \textbf{707} (2026), A176.

\bibitem{DiValentino:2021izs}
E.~Di Valentino \textit{et al.},
%``In the realm of the Hubble tension - a review of solutions,''
Class. Quantum Grav. \textbf{38} (2021), 153001.

\bibitem{Riess:2021jrx}
A.~G.~Riess \textit{et al.},
%``A Comprehensive Measurement of the Local Value of the Hubble Constant with 1% Uncertainty from Hubble Space Telescope Horizons,''
Astrophys. J. Lett. \textbf{934} (2022), L7.

\end{thebibliography}
\end{document}